\documentclass{ws-mpla}

\def\aa{{\it Astron. Astrophys.} \,}

\def\apj{{\it ApJ \,}}
\def\apjs{{\it Ap. J. Supp.} \,}

\def\apjl{{\it Ap. J. Lett.} \,}

\def\plb{{\it Phys. Lett. B.} \,}
\def\prd{{\it Phys. Rev. D.} \,}
\def\mn{{\it MNRAS} \,}

\def\na{{\it NewA} \,}

\def\ea{\ et al. \,}
\def\ngs{{non-Gaussian} \,}

\begin{document}

\markboth{Authors' Names}
{Instructions for Typing Manuscripts (Paper's Title)}

\catchline{}{}{}{}{}

\title{S-Z POWER SPECTRA}

\author{YOEL REPHAELI}

\address{School of Physics and Astronomy, Tel Aviv University\\
Tel Aviv, 69978, Israel\\
yoelr@wise.tau.ac.il}

\author{SHARON SADEH}

\address{School of Physics and Astronomy, Tel Aviv University\\
Tel Aviv, 69978, Israel\\
}

\maketitle

\pub{Received (Day Month Year)}{Revised (Day Month Year)}

\begin{abstract}

There is some observational evidence for earlier evolution of clusters 
of galaxies than predicted in the standard $\Lambda$CDM model with a 
Gaussian primordial density fluctuation field, and a low value for 
the mass variance parameter ($\sigma_8$). Particularly difficult in 
this model is the interpretation of possible excess CMB anisotropy on 
cluster scales as due to the Sunyaev-Zeldovich (S-Z) effect. We have 
calculated S-Z power spectra in the standard model, and in two alternative 
models which predict higher cluster abundance - a model with non-Gaussian 
PDF, and an early dark energy model. As anticipated, the levels of S-Z 
power in the latter two models are significantly higher than in the 
standard model, and in good agreement with current measurements of CMB 
anisotropy at high multipole values. Our results provide a sufficient 
basis for testing the viability of the three models by future high quality 
measurements of cluster abundance and the anisotropy induced by the S-Z 
effect.

\end{abstract}

\ccode{PACS Nos.: include PACS Nos.}

\section{Introduction}

The power spectrum of the primary CMB anisotropy falls sharply at 
mutlipoles $\ell>1000$, where `excess' power is primarily induced by 
the Sunyaev-Zeldovich (S-Z) effect. There are some indications that 
such excess power was measured by the CBI\cite{Readhead}, 
ACBAR\cite{Kuo}, and BIMA\cite{Dawson} experiments. Attributing the 
measured power to the S-Z effect requires a higher cluster abundance 
than predicted in the standard $\Lambda$CDM cosmological model, 
particularly so for the low value of the mass variance parameter,  
$\sigma_8=0.74^{+0.05}_{-0.06}$, deduced from the WMAP 3-year 
data\cite{Spergel}. An unrealistically high value $\sigma_8 \gtrsim 1$ 
would be required for consistency with the current CMB measurements. 
This possible discrepancy enhances interest in alternative models 
in which higher cluster abundances are predicted.

Other evidence for earlier and more abundant cluster population comes 
from radio observations of clusters at high redshift\cite{Miley,Venemans}, 
and larger values of the concentration parameter and Einstein radii than 
expected in the standard model\cite{Broadhurst}, which imply earlier 
formation of high mass ($M\geq 10^{15}\, M_{\odot}$) clusters. The 
cluster angular two-point correlation function provides yet another 
measure of the abundance and evolved nature of the population. A 
recent analysis of Spitzer Space Telescope observations
\cite{Magliocchetti} seems to indicate early formation of very massive 
galaxies, with number densities that are considerably higher than 
predicted in the standard $\Lambda$CDM model, and surprisingly high 
level of clustering.

The first of two alternative models, in which clusters are expected to 
form earlier than in the standard model, is an isocurvature CDM with 
scale-dependent non-Gaussian, $\chi^{2}_m$ distributed primordial density 
fluctuation field\cite{Peebles97,Peebles99}, where $m$ is the number of 
CDM fields added in quadratures to yield the $\chi^{2}_m$ distribution. 
With increased number of random primordial density fluctuation (PDF) 
fields, their sum approaches a Gaussian distribution (in accord with the 
central limit theorem); thus, the degree of deviation from a normally 
distributed PDF is the largest for $m=1$. The evolution of the large 
scale structure and primary CMB anisotropy in the $\chi^2$ family of 
models were explored in several studies\cite{Koyama,Mathis}, showing 
explicitly that primordial overdensities attain larger amplitudes with 
higher probabilities than in a Gaussian field. Correspondingly, cluster 
form earlier and are more abundant, thereby enhancing levels of S-Z 
observables\cite{Mathis,Sadeh1}. In the latter two papers the then 
current WMAP 1-year normalization was used, $\sigma_8=0.9$, and it was 
shown that already with this relatively high value it was difficult to 
reconcile the CMB power excess with the inferred cluster population, if 
the primordial fluctuation field was Gaussian. 

We note that the degree of non-Gaussianity in the explored $\chi^2$ 
models is consistent with limits set by analyses of the WMAP 
data\cite{Komatsu}. This is largely due to the fact that the latter 
dataset yields information on scales that are much larger than those 
associated with clusters. On these large scales the low overdensities 
may be indistinguishable from a non-Gaussian distribution when the 
density field is not scale invariant. 

Our previous work focused on the predicted levels of S-Z anisotropy and 
cluster number counts in the $\chi^{2}_1$ model\cite{Sadeh1,Sadeh2}, 
showing that S-Z power levels in this model are appreciably higher than 
indicated by current measurements. Here we explore predicted S-Z power 
spectra in the $\chi^{2}_2$ model.

Temporal variation of the dark energy density in early dark energy 
(EDE) models provides an alternative for generating an enhanced cluster 
abundance at higher redshifts. In these models the DE is appreciable 
already at early epochs and attains the observationally inferred value 
at present. The evolution of structure in the linear regime and CMB 
anisotropy in these models have been explored in several 
works\cite{Ferreira,Doran}. Two specific EDE models have recently been 
investigated in detail\cite{Bartelmann}, resulting in explicit numerical 
determination of the linear growth factor of density perturbations, 
critical density for spherical collapse, $\delta_c$, and overdensity at 
virialization, $\delta_v$, quantities needed to calculate the cluster 
mass function. The non-vanishing dark energy component at early times 
drives an early acceleration phase, implying a slower evolution of the 
linear growth factor and reduced values of $\delta_c$. Thus, for a given 
value of the mass variance normalization, $\sigma_{M}$, the corresponding 
quantity at early times should be larger than what is implied in the 
$\Lambda$CDM model, and since the critical overdensity for collapse at a 
given redshift is linearly extrapolated from an earlier time, the slower 
evolution of the growth factor in EDE models results in a lower 
$\delta_c$ as compared with its value in $\Lambda$CDM model. The 
reduced $\delta_c$ and higher $\sigma_{M}$ obviously yield a more 
abundant cluster population.

Here we update the predicted S-Z power spectra in the above two 
alternative cosmological models. In Section 2 we briefly describe 
the variant of the Press \& Schechter mass function in the 
$\chi^2$-distributed PDF, and outline the properties of the EDE model 
adopted in our calculations. Results of the calculations of S-Z 
power spectra in the $\Lambda$CDM, EDE, and non-Gaussian models 
are presented and compared in Section 3, followed by a brief 
discussion in Section 4.

\section{Calculations}

The calculations of S-Z power spectra requires knowledge and modeling of 
global, large scale, and cluster quantities. To do so we adopt the 
methodology described in several papers\cite{Komatsu2,Sadeh03}. We refer 
to the $\Lambda$CDM, EDE, and non-Gaussian models as models I, 
II, and III, respectively. Models I and III were described by 
us\cite{Sadeh1}, so our brief discussion here will include only the most 
essential aspects of these models. The EDE model we adopt here is 
characterized by the density parameter of early quintessence 
$\Omega_e=0.03$, and the coefficient of the equation of state parameter 
$w_0=-0.9$ at $z=0$. The effective coefficient as function of redshift 
is\cite{Wetterich}
\begin{equation}
\overline{w}(z)=\frac{w_0}{1+u\log{(1+z)}},
\end{equation}
where
\begin{equation}
u\equiv\frac{-3w_0}{\log{\left(\frac{1-\Omega_e}{\Omega_e}\right)}
+\log{\left(\frac{1-\Omega_m}{\Omega_m}\right)}},
\end{equation}
and $\Omega_{m}$ is the matter density parameter.

The basic quantity in our calculations of S-Z power spectra is the Press 
\& Schechter (1974) mass function,  
\begin{equation}
n(M,z)=-\mu F(\mu)\frac{\rho_b}{M\sigma_M^2(z)}\frac{d\sigma}{dM}\,dM,
\end{equation}
where $\mu\equiv\delta_c(z)/\sigma_M(z)$ is the ratio of the critical 
overdensity for collapse to the mass variance $\sigma_M$ at redshift $z$, 
and $\rho_b$ is the background density at $z=0$. For a Gaussian PDF field 
assumed in models I and II, we have 
\begin{equation}
F(\mu)=\sqrt{\frac{2}{\pi}}e^{-(\mu^2/2)}. 
\end{equation}
The corresponding expression for a PDF which obeys $\chi^2_2$ statistics 
of model III is particularly simple 
\begin{equation}
F(\mu)=e^{-(1+\mu)}.
\end{equation}
The mass function is normalized such that all the mass is included in 
halos, a normalization that in the original Press \& Schechter mass 
function was affected by including a (`fudge') factor of 2. The 
functional form of the mass function is different in model III from that 
in models I and II, and the redshift dependence of the critical density 
for collapse and the linear growth factor both differ in model II than 
the corresponding quantities in the other two models. Explicit 
expressions for these quantities are given in our recent 
paper\cite{Sadeh2}.

The mass variance $\sigma_M$ was calculated with a top-hat window
function and CDM transfer functions taken from Bardeen \ea 
(1986)\cite{Bardeen} - adiabatic transfer function for models I and 
II, and isocurvature transfer function for model III. The shape of the 
CDM transfer function in the EDE model is slightly different than in the 
standard model; this difference is ignored here. For the isocurvature 
model this function (of the wavenumber $k$) is 
\begin{equation}
T(k)_{isoc}=(5.6q)^{2}\left[1+\frac{(40q)^2}{1+215q+(16q)^2(1+0.5q)^{-1}}
+(5.6q)^{8/5}\right]^{-5/4},
\end{equation}
where $q\equiv k/(\Omega_mh^2Mpc^{-1})$, with 
$h=H_0/(100$ km s$^{-1}$ Mpc$^{-1}$). Values of the global parameters 
were taken to be those deduced from the WMAP 3-year data, 
$\Omega_{\Lambda}=0.76$ (which, for the case $w\ne -1$ 
is usually written as $\Omega_Q$), $\Omega_m=0.24$, $h=0.73$, and 
$\sigma_8=0.74^{+0.05}_{-0.06}$. The spectral index of the PDF spectrum 
is $n=1$ in models I and II, and $n=-1.8$ in model III. In the EDE model 
the differential equations for the evolution of the linear growth factor, 
$\delta_c$, and $\delta_v$, are different than in the other two models. 
Full description of the of these equations and their numerical solutions 
can be found in our recent paper\cite{Sadeh2}. 

A meaningful comparison of the predicted S-Z spectra of the three 
models considered here requires proper and self-consistent 
normalization of the respective mass functions. This is accomplished 
by requiring that the cumulative cluster density at $z=0$ is in 
agreement with that calculated in the standard $\Lambda$CDM model.

In addition to the global and large-scale parameters, the calculation 
of S-Z power spectra necessitates full description of the properties of 
IC gas. To do so clusters are assumed to be in hydrostatic equilibrium 
with $\beta$ profiles for the total mass and gas density profiles, and 
with a gas mass fraction of $0.1$. The gas temperature is determined 
from the equation of hydrostatic equilibrium.

\section{Sample Power Spectra}

\begin{figure}
\centering
\epsfig{file=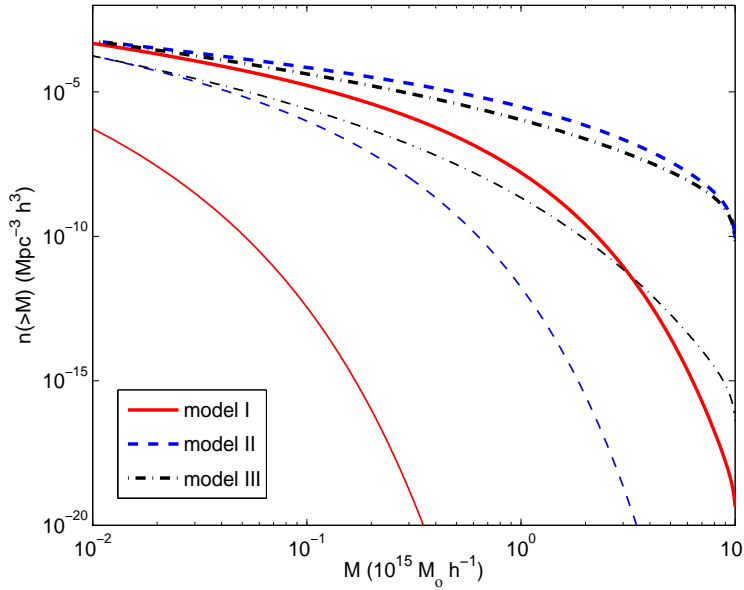, height=8.5cm, width=11cm, clip=}
\caption{Cumulative mass functions in the three models. The upper and lower 
curves for each model correspond to redshifts $z=0.01$ and $z=3$, 
respectively.}
\label{fig:mf}
\end{figure}

The behavior of the cumulative mass function in the three models 
at $z=3$ and $z=0.01$ is shown in Fig.~\ref{fig:mf} for the mass range 
$10^{13}M_{\odot}h^{-1}\le M\le 10^{16}M_{\odot}h^{-1}$. As is clear, 
the models are correctly normalized to yield the same cumulative cluster 
density at low redshifts. Abundances of high-mass clusters are indeed 
higher in models II and III than in model I. As anticipated, the relative 
abundances with respect to those in $\Lambda$CDM increase with redshift. 
It is also apparent that at early times the excess of high overdensity 
fluctuations in the $\chi^{2}_2$ model has a stronger impact on the 
abundance of high mass clusters than the slower evolution of the linear 
growth factor and lower values of $\delta_c$ in model II, whereas their 
respective effects at present are about the same in these two models.

\begin{figure}
\centering
\epsfig{file=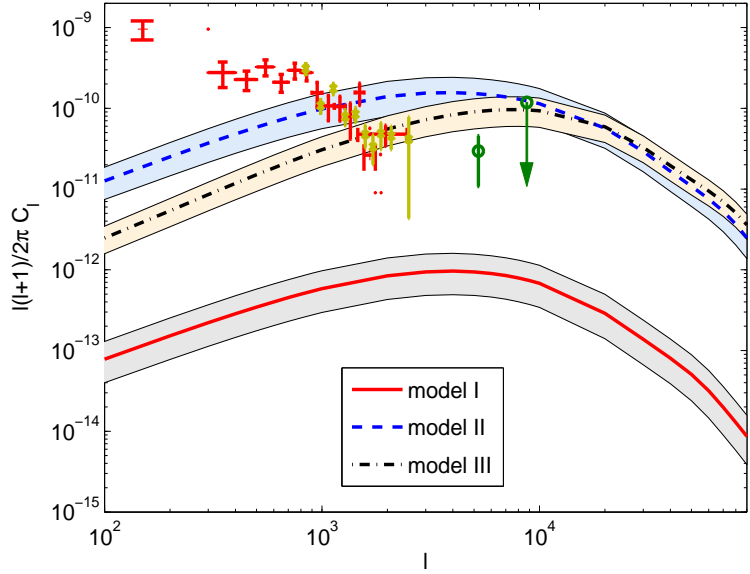, height=8.5cm, width=11cm, clip=}
\caption{S-Z angular power spectrum at $\nu=31\,GHz$ in 
models I (continuous), II (dashed), III (dash-dotted), and IV (thick 
dash-dotted). The shaded areas correspond to the WMAP reported 1-$\sigma$ 
errors in $\sigma_8$. Also shown is the power excess reported by the BIMA 
(diamonds), CBI (crosses) and ACBAR (x-symbols) experiments. Note that the 
BIMA indicated power at $\ell=8748$ is an upper limit.}
\label{fig:ps}
\end{figure}

The enhanced cluster abundances in models II and III are directly manifested 
in higher S-Z power levels than in the standard model, as is immediately 
evident in Fig.~\ref{fig:ps} which shows the spectra at 31 GHz together with 
CBI, ACBAR, and BIMA measurements at this frequency. In addition, the (broad) 
peak power in the \ngs model is reached at multipoles, $\ell \sim 7000-8000$ as 
compared with $\ell \sim 4000-5000$ in model I and II, a direct consequence 
of the higher abundances of distant clusters (with smaller apparent sizes) 
in model III. 

The predictions of both the EDE and \ngs models are more consistent with 
current measurements of the CMB power spectrum. The shape of their S-Z 
spectra are virtually identical for $\ell\geq 10^4$, with power levels 
somewhat lower at lower $\ell$ in the former model than in the \ngs model.

\section{DISCUSSION}

The work reviewed here has been motivated by initial indications that 
there might be a significant discrepancy between current measurements of 
levels of CMB anisotropy on scales $\ell \geq 2000$, and predicted levels of 
S-Z power in the standard $\Lambda$CDM with gaussian PDF field. The 
discrepancy stems mainly from the fact that high-mass clusters, the largest 
contributors to S-Z power whose density decreases sharply with decreasing 
$\sigma_8$, are not sufficiently abundant if this important parameter is 
as low as deduced from the 3-year WMAP data. A higher cluster abundance 
is expected when the PDF field has the form predicted in the non-gaussian 
$\chi^2_2$ model. This is a result of higher probabilities for overdense 
regions at high $z$, leading to earlier collapse of proto-cluster halos. 
In the early quintessence model earlier cluster collapse is a manifestation 
of higher linear growth factor, and lower value of the critical density for 
collapse at high $z$. Accordingly, levels of S-Z power are higher in these 
two alternative models. Clearly, these two non-standard models are by no 
means the only viable alternatives to the standard model. As we noted, the 
non-Gaussian model considered here is just one of the $\chi^2_m$ family, but 
with decreasing non-Gaussianity with increasing $m$. Also, other early 
quintessence models with higher EDE densities result in slower evolution of 
the linear growth factor, and reduced value of $\delta_c$, thereby leading 
to higher levels of S-Z power.

While levels of S-Z power spectra predicted in the standard model span an 
appreciable range, reflecting also uncertainties in the evolution of internal 
properties of clusters - such as IC gas density and temperature - their 
maximal values are still well below current observational results. This 
conclusion is based on extensive investigations of the cluster 
temperature-mass relations\cite{Sadeh03,Sadeh04}, IC gas models, including 
non-isothermal polytropic temperature profiles\cite{Komatsu2,Sadeh1}, and 
evolution of the gas mass fraction.

Our purpose here has been to show that the two alternative models explored 
can produce S-Z power levels that are substantially higher than in the 
standard model. It is too early to actually fit the predictions to the 
preliminary high $\ell$ results. This will have to be done in conjunction 
with other cluster observables, such as cluster (S-Z) number counts, and 
the two-point correlation function (which were considered in our previous 
work\cite{Sadeh2}). Other cluster measures that are very much affected in 
these models are formation times and concentration parameters (whose 
observational manifestations include, e.g., mass profiles and Einstein 
ring sizes). Nonetheless, it is quite clear from our results that the 
$\chi^2_{1}$ model, and EDE models with larger quintessence densities at early 
times than in the specific model adopted here, predict significantly higher 
levels of S-Z power than indicated by current observational results, and 
therefore seem to be non-viable. 

We should know soon whether the apparent discrepancy is real, when results 
of more extensive and precise measurements of the high $\ell$ power will 
become known. But irrespective of these upcoming CMB and S-Z measurements, 
there seem to be other observational indications that cluster formed earlier 
and are more abundant than predicted in standard $\Lambda$CDM. The 
non-gaussian and EDE models explored here seem to be viable alternatives, 
should the discrepancy persist.

\section{ACKNOWLEDGMENT}

Work at Tel Aviv University is supported by a grant from the Israel Science 
Foundation.

\end{document}